\documentclass[aps,prb,superscriptaddress,twocolumn]{revtex4-1}

\usepackage{graphics}
\usepackage{graphicx} 
\usepackage{dcolumn}
\usepackage{bm}

\begin{document}


\title{Magnetic structure and spin excitations in BaMn$_2$Bi$_2$}



\author{S.~Calder}
\email{caldersa@ornl.gov}
\affiliation{Quantum Condensed Matter Division, Oak Ridge National Laboratory, Oak Ridge, TN 37831.}

\author{B.~Saparov}
\affiliation{Materials Science and Technology Division, Oak Ridge National Laboratory, Oak Ridge, TN 37831.}

\author{H.~B.~Cao}
\affiliation{Instrument and Source Division, Oak Ridge National Laboratory, Oak Ridge, TN 37831.}

\author{J.~L.~Niedziela}
\affiliation{Quantum Condensed Matter Division, Oak Ridge National Laboratory, Oak Ridge, TN 37831.}

\author{M.~D.~Lumsden}
\affiliation{Quantum Condensed Matter Division, Oak Ridge National Laboratory, Oak Ridge, TN 37831.}

\author{A.~S.~Sefat}
\affiliation{Materials Science and Technology Division, Oak Ridge National Laboratory, Oak Ridge, TN 37831.}
 
\author{A.~D.~Christianson}
\affiliation{Quantum Condensed Matter Division, Oak Ridge National Laboratory, Oak Ridge, TN 37831.}


\date{\today}

\begin{abstract}
We present a single crystal neutron scattering study of BaMn$_2$Bi$_2$, a recently synthesized material with the same ThCr$_2$Si$_2$-type structure found in several Fe-based unconventional superconducting materials. We show long range magnetic order, in the form of a G-type antiferromagnetic structure, to exist up to 390 K with an indication of a structural transition at 100 K. Utilizing inelastic neutron scattering we observe a spin-gap of 16meV, with spin-waves extending up to 55 meV. We find these magnetic excitations to be well fit to a $J_1$-$J_2$-$J_c$ Heisenberg model  and  present values for the exchange interactions. The spin wave spectrum appears to be unchanged by the 100 K structural phase transition.
 \end{abstract}


\maketitle

\section{Introduction}

The  discovery of unconventional superconductivity in Fe-based materials has stimulated intense interest in the condensed matter physics community, and offers a new system, with the same square planar crystal motif as the cuprates, in which to investigate the underlying mechanism of unconventional superconductivity \cite{KamiharaFebased,JohnstonReview,LumsdenSCreview}. Driven by the initial discovery of superconductivity several new Fe-based materials have been uncovered, for example the 1111 phase ($R$FeAsO with $R$ = rare earth), 111 phase ($A$FeAs with $A$ = alkali metal), 11 phase (FeTe or FeSe), 122 phase ($A$Fe$_2$As$_2$), and FeSe122 phase ($A$Fe$_2$Se$_2$) \cite{LumsdenSCreview, JohnstonReview}. The FeAs (or FeSe) layers are the common ingredient and unlike the cuprates it is possible to dope the Fe site and attain superconductivity. As a consequence much work has been done on this, particularly in the 122 phase. 

Recent interest has also focused on complete substitution of the Fe ion in the 122-type structure as an avenue to both search for new classes of superconducting materials or probe why no superconductivity is attained despite often similar physics. One pertinent example being SrCo$_2$As$_2$ that shows many similar features to the Fe-based 122 materials, but as yet no superconductivity has been uncovered upon doping from the cobalt parent side  \cite{PhysRevLett.111.157001}. Additionally the chromium based material BaCr$_2$As$_2$ was shown to host itinerant antiferromagnetism that differs from the Fe-122 materials that remains upon doping and prohibits superconductivity \cite{PhysRevB.79.094429, PhysRevB.83.060509}.  

Directly related to our investigation is the Mn-122 material BaMn$_2$As$_2$ that shows alternative behavior to both Co and Fe-122 materials as well as the cuprates.   BaMn$_2$As$_2$ is a G-type antiferromagnet (AFM),  T$\rm _N$=625 K, with no structural transition in the magnetic phase \cite{PhysRevB.80.100403, PhysRevB.84.094445}. Investigations of BaMn$_2$As$_2$ have indicated properties in the  intermediate regime  between those of the itinerant AFe$_2$As$_2$ antiferromagnets and the local moment antiferromagnetic insulator La$_2$CuO$_4$ parent superconductors. Indeed doping BaMn$_2$As$_2$ in the form Ba$_{1-x}$K$_x$Mn$_2$As$_2$ has been shown to result in an antiferromagnetic local moment metal \cite{PhysRevLett.108.087005}. Therefore it has been suggested that Mn-122 compounds may be well placed to act as a bridge between Fe and Cu based unconventional superconductors \cite{JohnstonReview}.

Ref.~\onlinecite{Saparov201332} reported the growth and characterization of single crystals of the new Mn-122 material, BaMn$_2$Bi$_2$ \cite{Saparov201332} the first bismuthide with ThCr$_2$Si$_2$-type structure (space group $I4/mmm$ with a= 4.4902(3) $\rm \AA$ and c=14.687(1) $\rm \AA$). BaMn$_2$Bi$_2$ is insulating with a small band gap of E$_g$ = 6 meV, with metallic behavior achieved via hole doping with 10$\%$ K substitution on the Ba site \cite{Saparov201332}. Susceptibility measurements revealed an anomaly around 400 K consistent with magnetic ordering, with an additional apparent anomaly around 100 K that also corresponds to an upturn in the resistivity \cite{Saparov201332}.

BaMn$_2$Bi$_2$ shows similar properties to BaMn$_2$As$_2$ and therefore may be similarly placed to act as a bridge between Fe and Cu based unconventional superconductors. Additionally BaMn$_2$Bi$_2$ offers an alternative to arsenic based materials and is amenable to the growth of suitably large single crystals for neutron scattering. Here we report the results of a single crystal neutron diffraction investigation of  BaMn$_2$Bi$_2$ through both the 400 K and 100 K anomalies observed from bulk measurements.  We then extend our exploration of the physical properties of BaMn$_2$Bi$_2$ to a study of the spin excitation spectrum by means of inelastic neutron scattering measurements. Our results support the postulate that the Mn-122 series hosts magnetic and insulating  properties intermediate between Fe and Cu based materials and we utilize our single crystal inelastic neutron scattering results to provide detailed information on the magnetic exchange interactions and spin gap.

\section{Experimental Methods}

A detailed description of the single crystal growth and characterization of BaMn$_2$Bi$_2$  is presented in Ref.~\onlinecite{Saparov201332}. Single crystal neutron scattering measurements were performed on the Four-circle diffractometer (HB-3A) at the High Flux Isotope Reactor at ORNL. A single crystal of  $\sim$100 mg was measured in the temperature range 4 K to 400 K and the data refined using Fullprof to obtain crystal and magnetic structures. To attain a suitable mass ($\sim$3 grams) for inelastic neutron scattering measurements five single crystals were coaligned to within 1$^\circ$  in the ($HHL$) scattering plane.  Utilizing the ARCS spectrometer at the SNS, ORNL, inelastic neutron measurements were performed at 4 K and 120 K with incident energies of 60, 100, 120, 250 and 500 meV. To examine the spin excitations in all reciprocal lattice directions the sample was rotated by 90$^\circ$ in 1$^\circ$ for measurements with an incident energy of 100 meV. The different angular data was combined and subsequent cuts performed with the Horace software \cite{Horace}. The instrument resolution varies with energy transfer and this was accounted for  in our fitting of the data through the use of an analytical function described in Ref.~\onlinecite{abernathy:015114}. The  inelastic energy resolution at an energy transfer of 50 meV is 1.56 meV for 100 meV incident energy. 

\section{Results and Discussion}

\subsection{Magnetic and nuclear structure of BaMn$_2$Bi$_2$}


\begin{figure}[t]
   \centering                   
 \includegraphics[trim=0.0cm 0.0cm 0.0cm 0.0cm,clip=true, width=1.0\columnwidth]{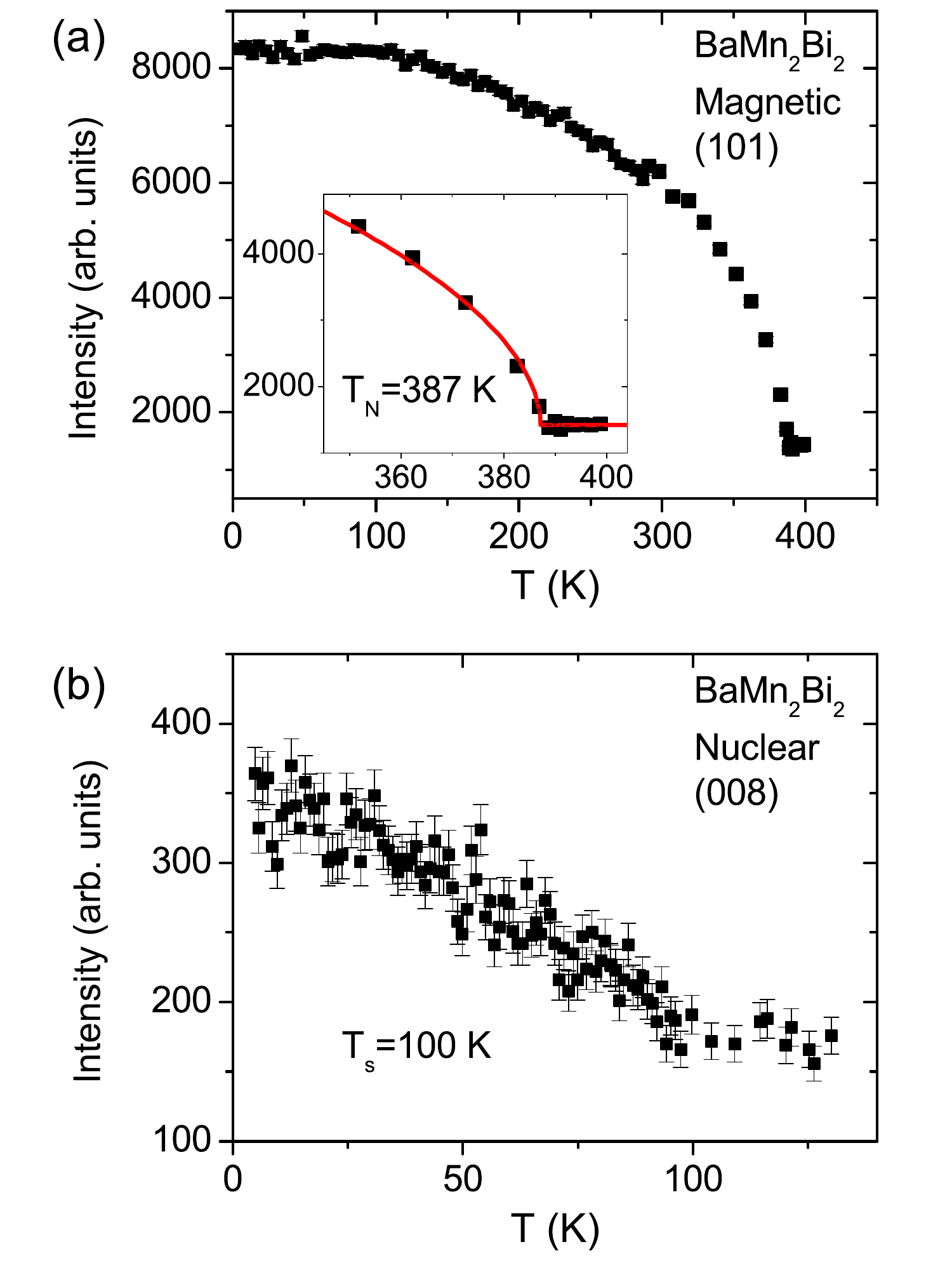} 
                   \caption{\label{Mag_OP} Single crystal neutron scattering measurements on BaMn$_2$Bi$_2$. (a) Magnetic ordering occurs at (101) reflections, with the onset at 387.2(4) K. (b) A subtle structural transition is indicated by the change in the nuclear (008) reflection below 100 K.}
\end{figure}

We begin our investigation of BaMn$_2$Bi$_2$  with single crystal Four-circle neutron diffraction measurements. We performed measurements on several different nuclear reflections and found the refined structure to be consistent with the previously reported powder x-ray measurements \cite{Saparov201332}. From susceptibility measurements there is an anomaly around 400 K, attributed magnetic ordering \cite{Saparov201332}. We observed intensity at the (101) reflection, a reflection forbidden by the nuclear space group symmetry but consistent with AFM  long range order. Indeed this defines the propagation vector for BaMn$_2$Bi$_2$ with space group $I4/mmm$ in the body centered tetragonal notation, in primitive tetragonal notation the propagation vector transforms to ($\frac{1}{2}$,$\frac{1}{2}$,$\frac{1}{2}$). Following the temperature evolution of the intensity at the (101) reflection position from 4 K to 400 K we observe a magnetic transition at  T$\rm _N=$387.2(4) K, shown in Fig.~\ref{Mag_OP}. Considering the intensity of magnetic reflections in different Brillouin zones at 4 K within a model based on equal populations of domains for tetragonal symmetry, and normalizing to the nuclear reflections, we were able to define the magnetic structure in BaMn$_2$Bi$_2$ as being G-type AFM, with the spins aligned along the $c$-axis. The ordered moment at 4 K is 3.83(4)$\rm \mu_B$/Mn, reduced from the 5$\rm \mu_B$/Mn expected for the high-spin S=5/2 of Mn$^{2+}$, nevertheless closer to a local moment description relative to the itinerant Fe-based superconductors. The magnetic structure, and within error the ordered moment, is the same as for the related Mn-122 BaMn$_2$As$_2$  \cite{PhysRevB.80.100403}.  

\begin{figure}[tb]
   \centering                   
                  \includegraphics[trim=2.5cm 0.5cm 2.5cm 0.5cm, clip=true, width=0.9\columnwidth]{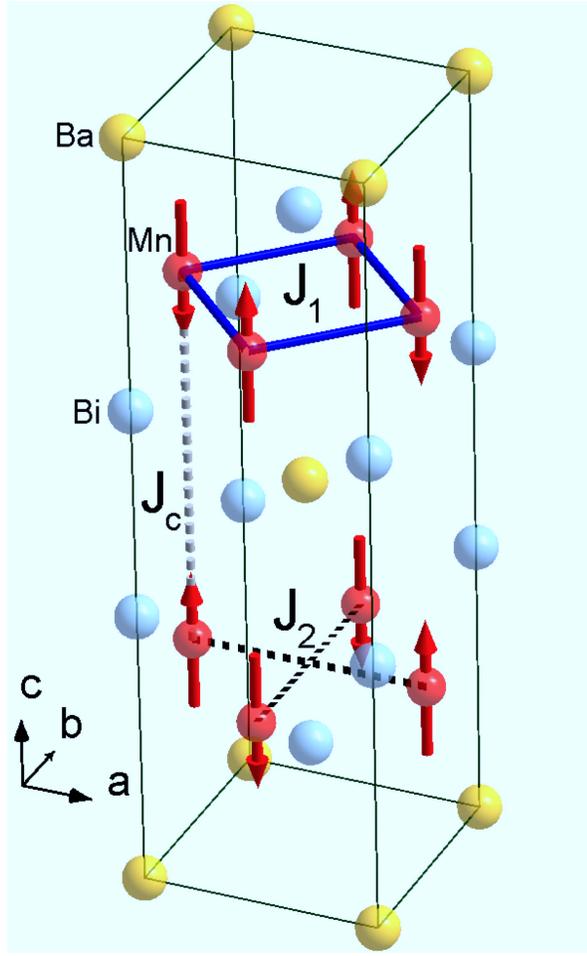} 
                   \caption{\label{xtal_Js}  Crystal structure of BaMn$_2$Bi$_2$ with  G-type AFM ordering of the Mn-ions shown within  the nuclear unit cell.  The exchange interactions used in the model Hamiltonian correspond to $J_1$ (ab-plane nearest neighbor), $J_2$ (ab-plane next nearest neighbor) and $J_c$ ($c$-axis nearest neighbor).}                
\end{figure}

Both susceptibility and resistivity measurements on BaMn$_2$Bi$_2$ suggest a further transition around 100 K \cite{Saparov201332}. The behavior of the (101) magnetic reflection in Fig.~\ref{Mag_OP}(a), however, shows no observable change in the magnetic structure in this region. Fig.~\ref{Mag_OP}(b) shows the nuclear (008) reflection, with an anomaly at the same temperature as that observed in bulk measurements. Such a change is indicative of a subtle tetragonal to orthorhombic transition that is not observable as a peak splitting due to the instrument resolution, but would be manifested in a change in the extinction and therefore the measured intensity of a nuclear reflection. A tetragonal to orthorhombic structural change is observed in several Fe-122 parent superconductors and is perhaps a prerequisite for attaining superconductivity, however no structural change in the magnetic ordered phase has been observed in the related BaMn$_2$As$_2$.  


\subsection{Spin excitations of BaMn$_2$Bi$_2$}

\begin{figure}[h]
   \centering                   
     \includegraphics[trim=0.7cm 6.2cm 0.7cm 6.2cm,clip=true, width=0.49\columnwidth]{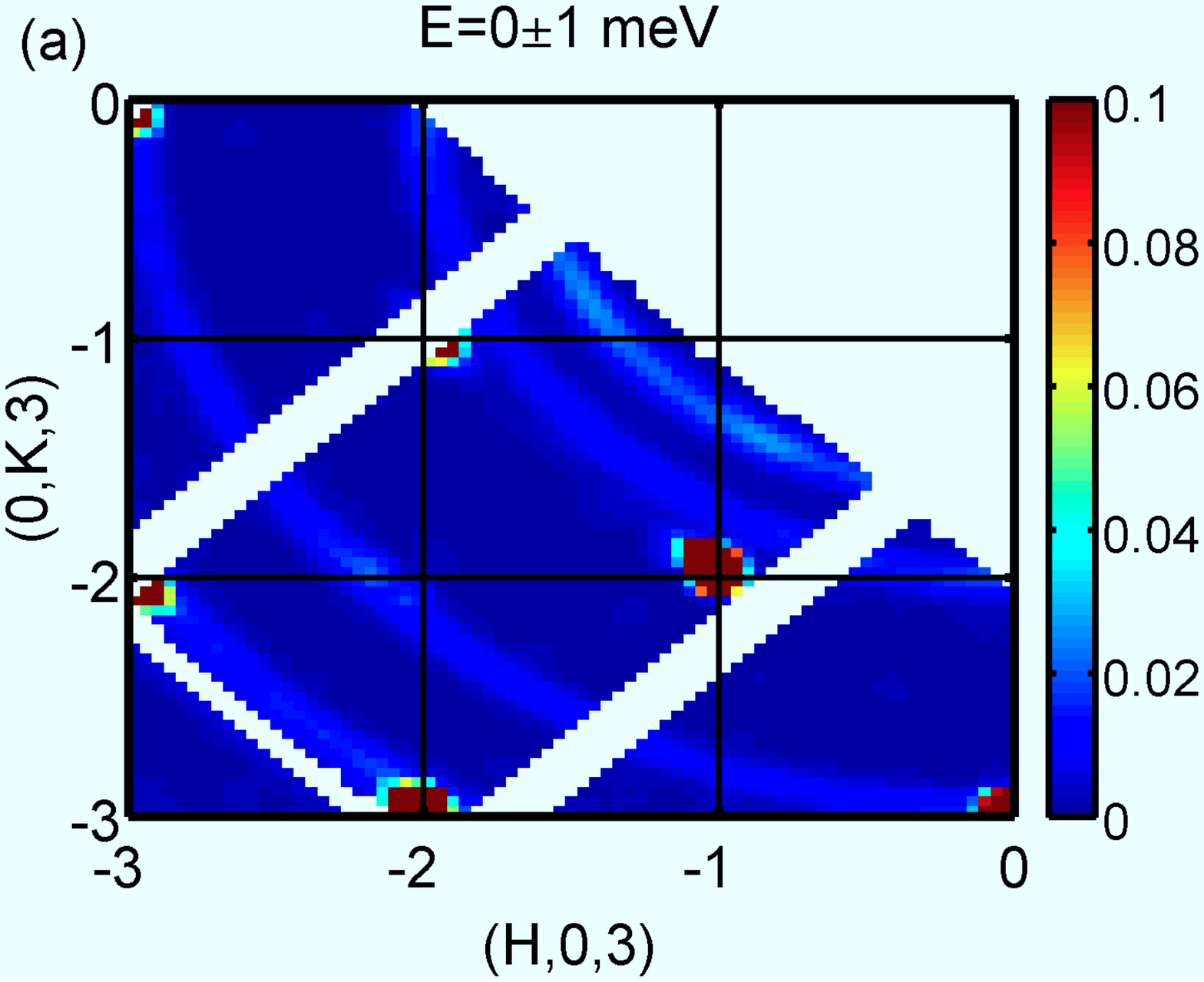} 
           \includegraphics[trim=0.7cm 6.2cm 0.7cm 6.2cm,clip=true, width=0.49\columnwidth]{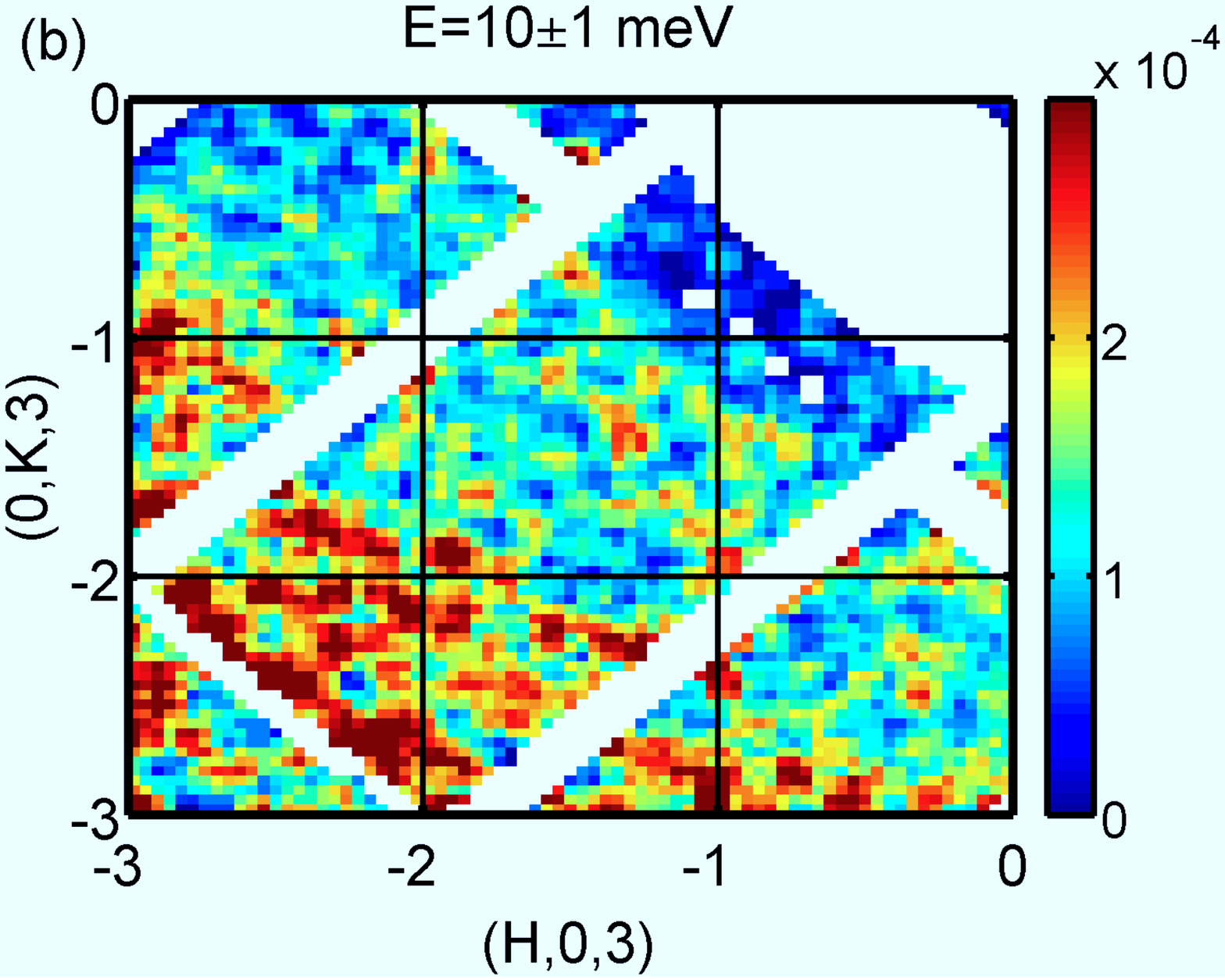} 
              \includegraphics[trim=0.7cm 6.2cm 0.7cm 6.2cm,clip=true, width=0.49\columnwidth]{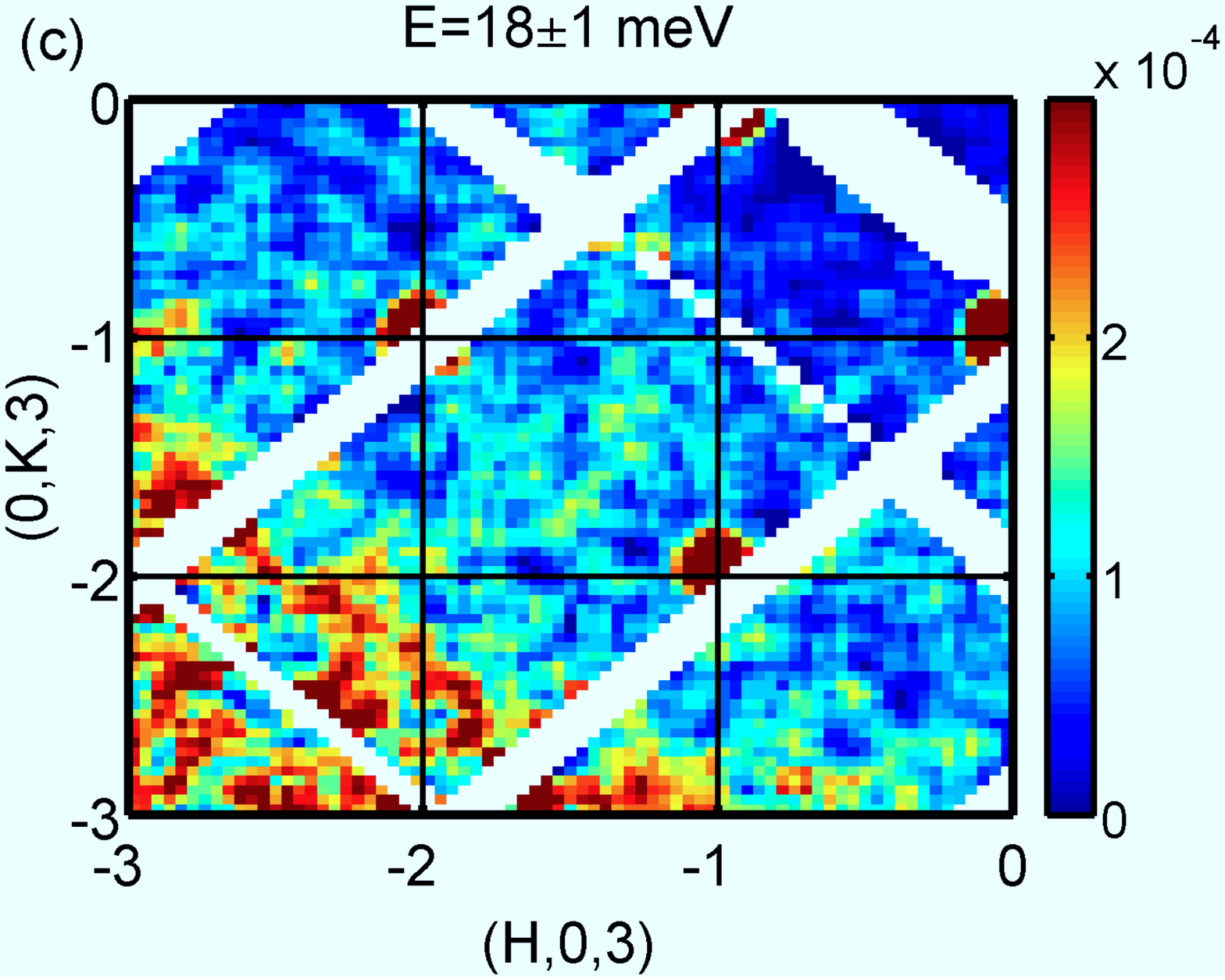} 
                 \includegraphics[trim=0.7cm 6.2cm 0.7cm 6.2cm,clip=true, width=0.49\columnwidth]{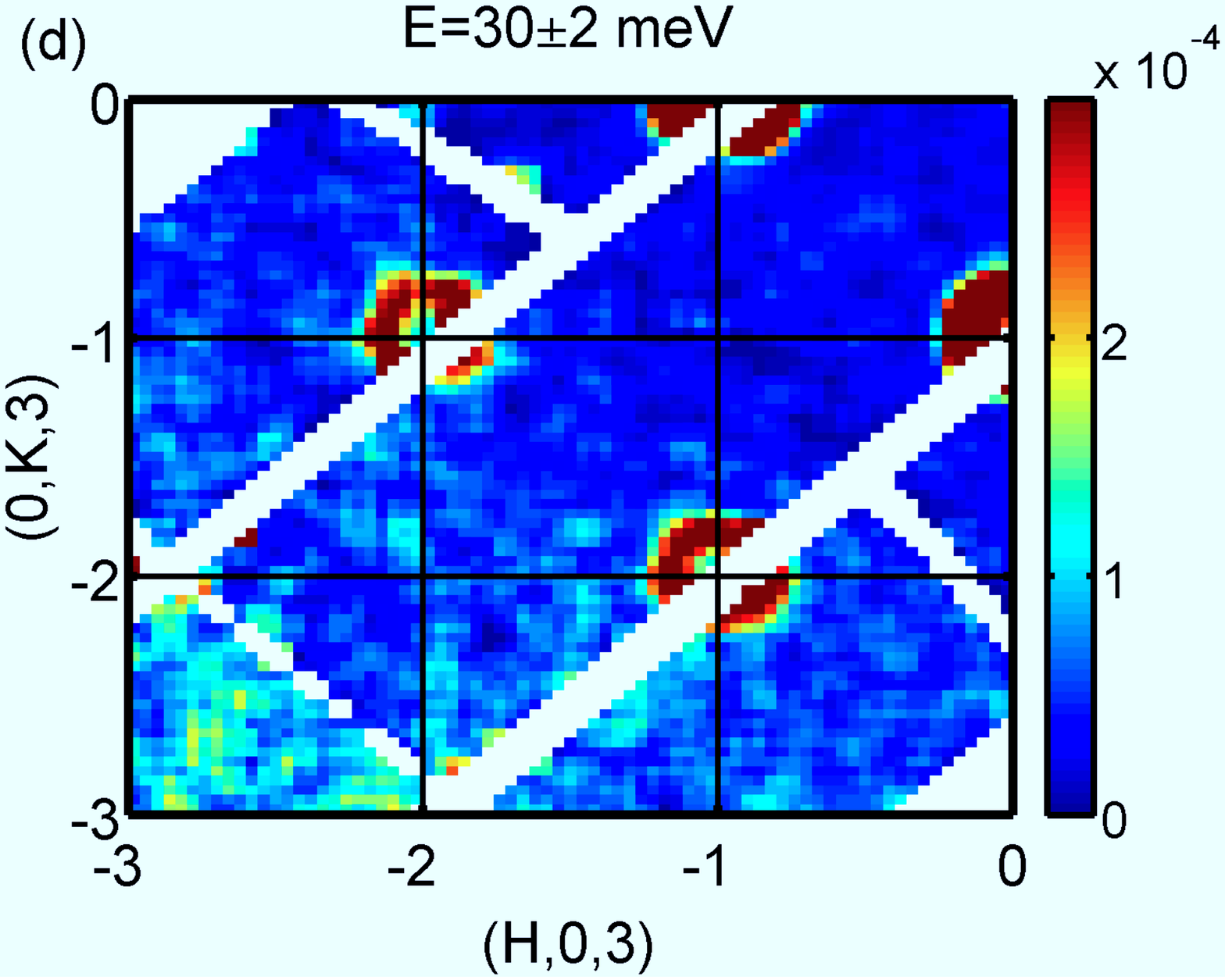} 
                    \includegraphics[trim=0.7cm 6.2cm 0.7cm 6.2cm,clip=true, width=0.49\columnwidth]{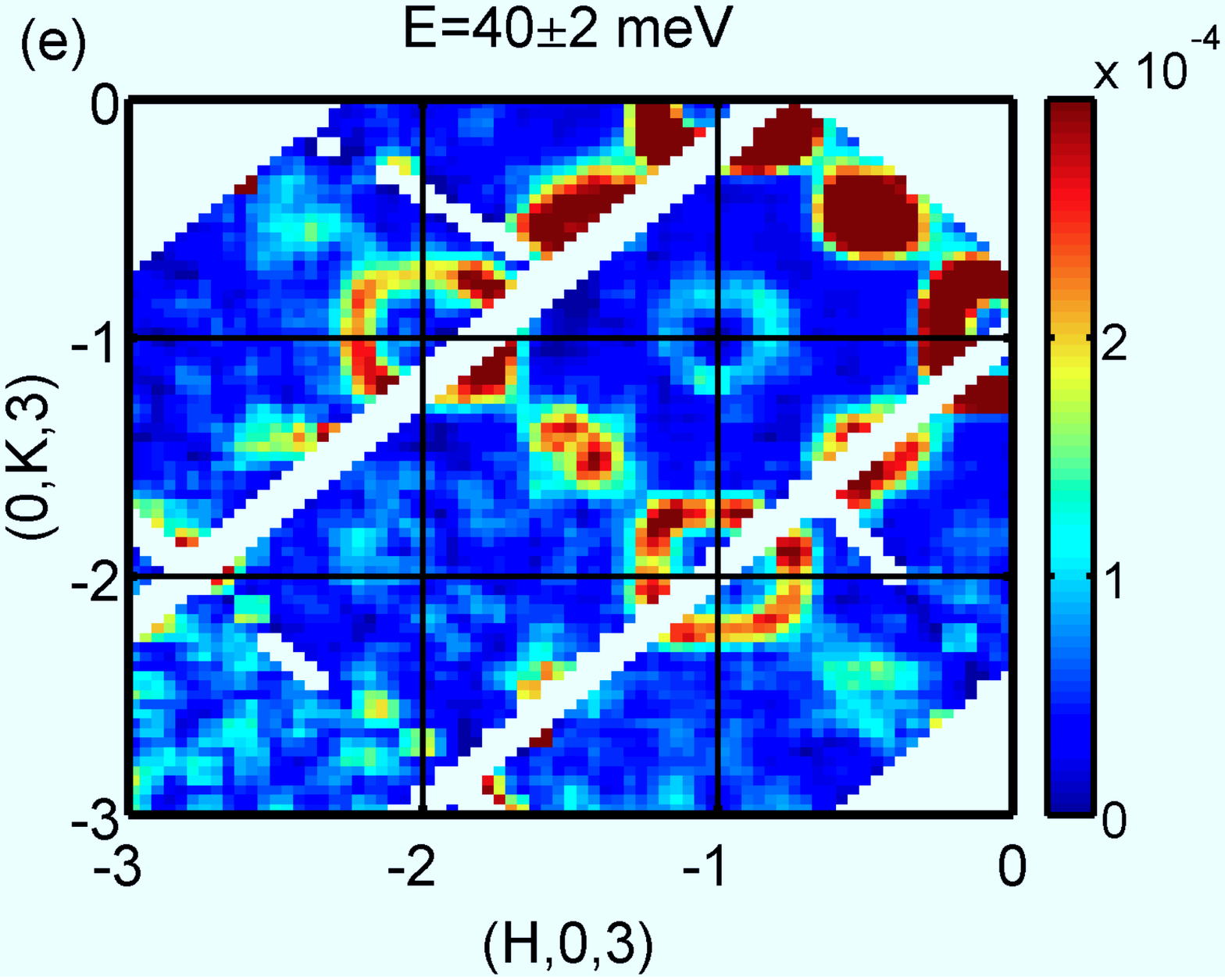} 
                       \includegraphics[trim=0.7cm 6.2cm 0.7cm 6.2cm,clip=true, width=0.49\columnwidth]{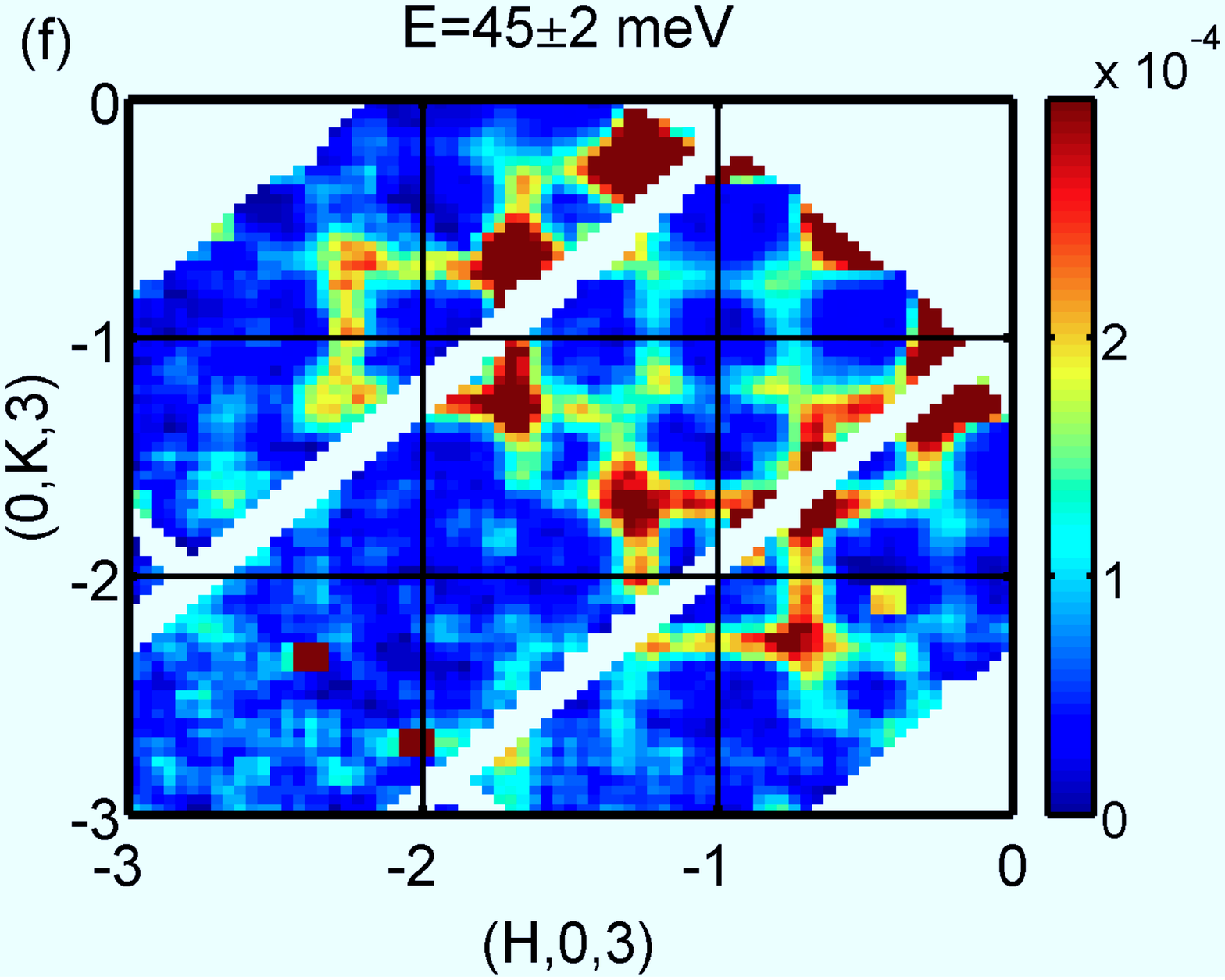} 
                          \includegraphics[trim=0.7cm 6.2cm 0.7cm 6.2cm,clip=true, width=0.49\columnwidth]{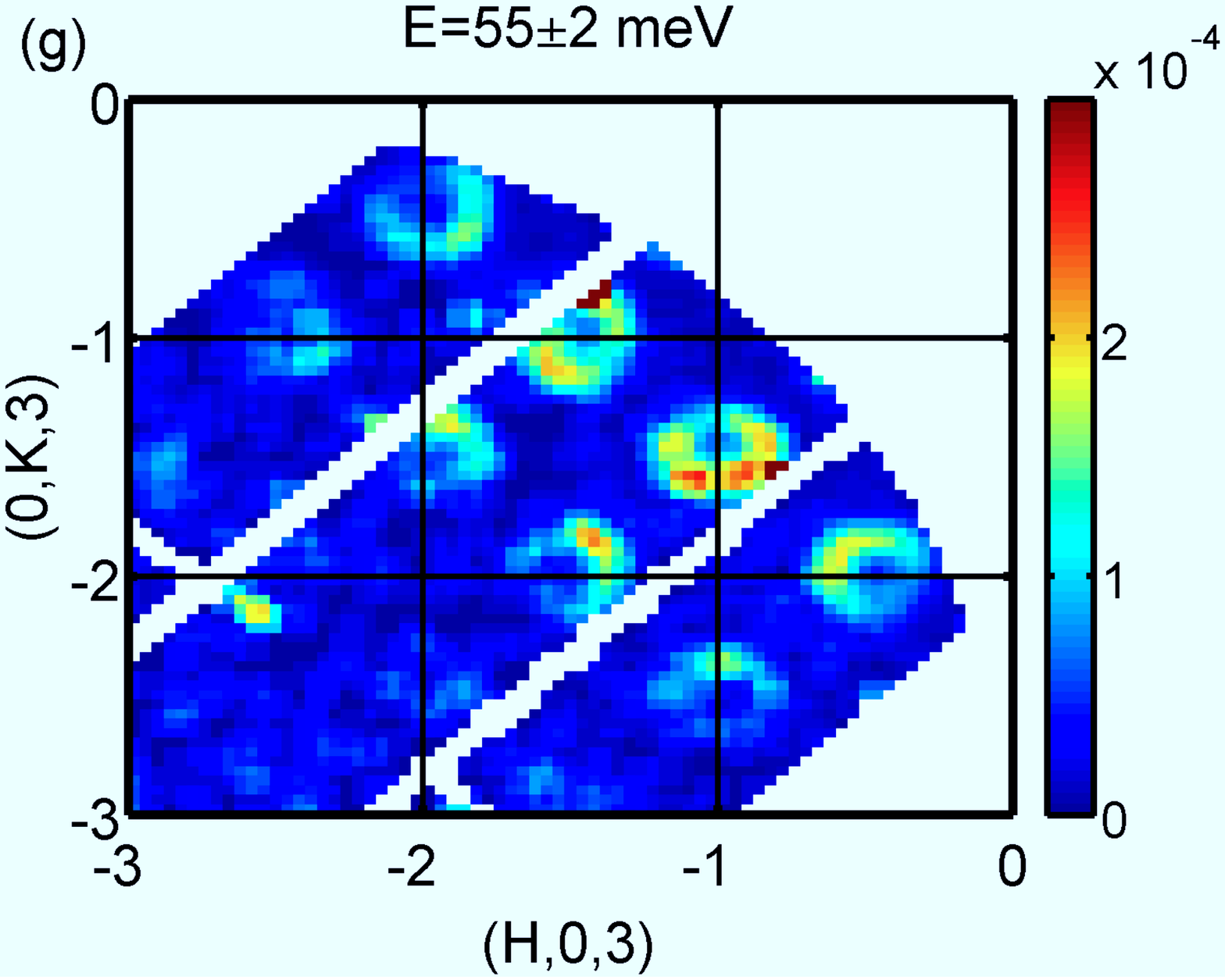} 
                             \includegraphics[trim=0.7cm 6.2cm 0.7cm 6.2cm,clip=true, width=0.49\columnwidth]{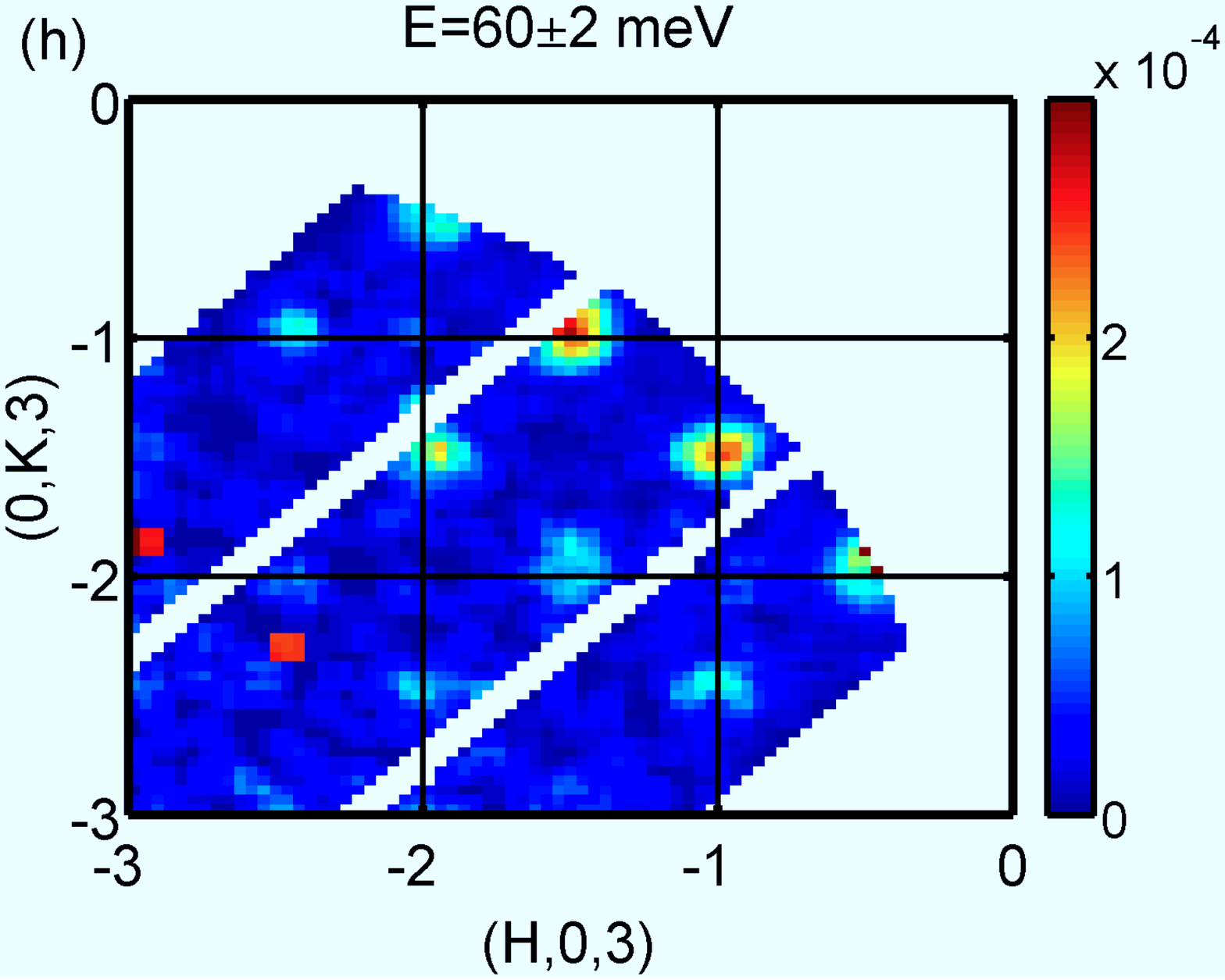} 
                                  \includegraphics[trim=0.0cm 0cm 0cm 0cm,clip=true, width=0.9\columnwidth]{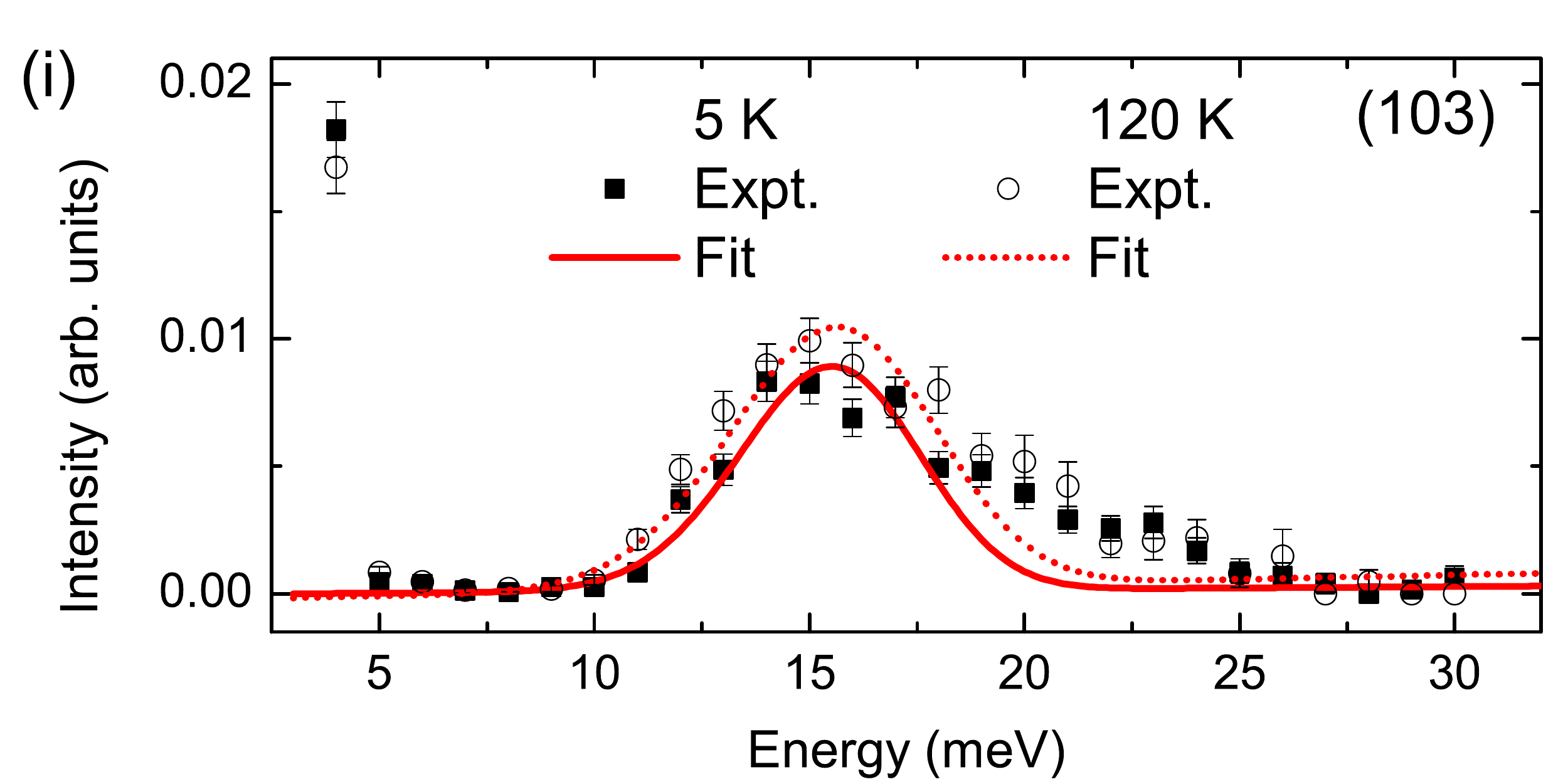} 
                   \caption{\label{ConstEcuts} Inelastic neutron scattering measurements on single crystals of BaMn$_2$Bi$_2$ with an incident energy of 100 meV.~(a)-(h) Constant energy cuts projected onto the (H,K) plane for fixed L=3. The scale corresponds to intensity in arbitrary units. Excluding the E=0 meV slice the scale remains fixed for each plot.~(i) Constant q=(1,0,3) cut from the magnetic Bragg reflection in the range (0.93$<$H$<$1.07, -0.07$<$K$<$0.07, 2.93$<$L$<$3.07) to reveal the spin-gap. The fit is performed excluding the region around 20 meV to remove background.}
\end{figure}

We now consider the magnetic excitations through a single crystal inelastic neutron scattering experiment on ARCS with the aim of finding the exchange interactions and observing any spin-gap,  in addition to comparing excitations through the apparent 100 K structural transition. We performed an initial survey using incident energies of E$\rm _i$=60, 120, 250 and 500 meV with the incident beam along the $c$-axis. This allowed us to find the top of the spin excitations at $\sim$60 meV and rule out any higher energy spin waves. With the spin excitations residing around 60 meV and below we chose an E$\rm _i$=100 meV to map out the low Q Brillouin zones and performed cuts along high symmetry directions. Constant energy cuts are shown in Fig.~\ref{ConstEcuts}(a)-(h) that follow the evolution of the low energy excitations.

\clearpage

Familiar spin-wave cones are seen to develop centered on the magnetic Bragg points (Fig.~\ref{ConstEcuts}(a)-(d)), with overlapping excitations near the maximum of the branches (Fig.~\ref{ConstEcuts}(e)-(f)). The lack of any Q-dependent scattering in Fig.~\ref{ConstEcuts}(b) at 10 meV indicates the existence of a spin-gap in BaMn$_2$Bi$_2$. This is confirmed by an energy cut from the elastic magnetic Bragg reflection position (103), as shown in Fig.~\ref{ConstEcuts}(i) where a distinct energy gap is observed well within the instrumental energy resolution of 2.7 meV at 5 meV energy transfer. No change in the gap is apparent through  100 K, where we observed a subtle structural transition from our single crystal neutron diffraction. In order to fit the data we exclude the region around 20 meV where background aluminum scattering occurs and fit the resulting profile to a gaussian convoluted with the instrument resolution to give an energy gap of E$_g = 16.29(26)$ meV at 5 K. The existence of a spin gap between $E_g$ = 6 to 9 meV is a general feature of the inelastic neutron spectrum of parent Fe-122 materials \cite{PhysRevLett.101.227205, PhysRevLett.101.167203}, with debate existing as to the importance and consequence on the emergence of superconductivity. No spin-gap is reported from the inelastic neutron scattering results of BaMn$_2$As$_2$ \cite{PhysRevB.84.094445}. This may be a consequence of only polycrystalline BaMn$_2$As$_2$ being synthesized or conversely point to different physics between the Mn-122 materials.

\begin{figure}[tb]
   \centering                   
     \includegraphics[trim=0.7cm 1cm 0.7cm 1cm,clip=true, width=1.0\columnwidth]{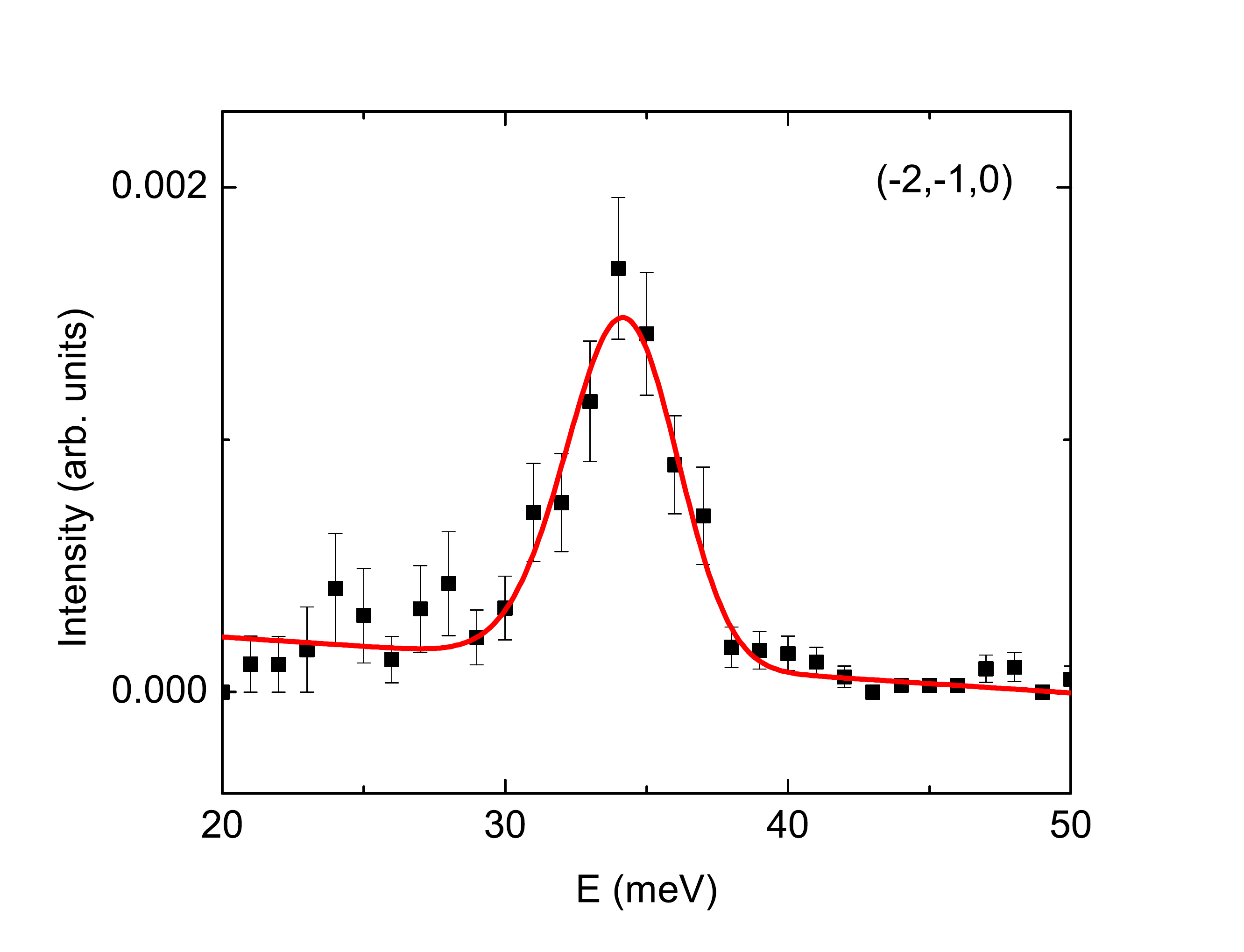} 
                   \caption{\label{ZB} Energy variation showing the width of the excitation through the zone boundary at (-2,-1,0) at 4 K. The fit is a gaussian convolved with the instrument energy resolution.}
\end{figure}

\begin{figure}[tb]
   \centering                   
     \includegraphics[trim=0.7cm 6.2cm 0.7cm 6.2cm,clip=true, width=1.0\columnwidth]{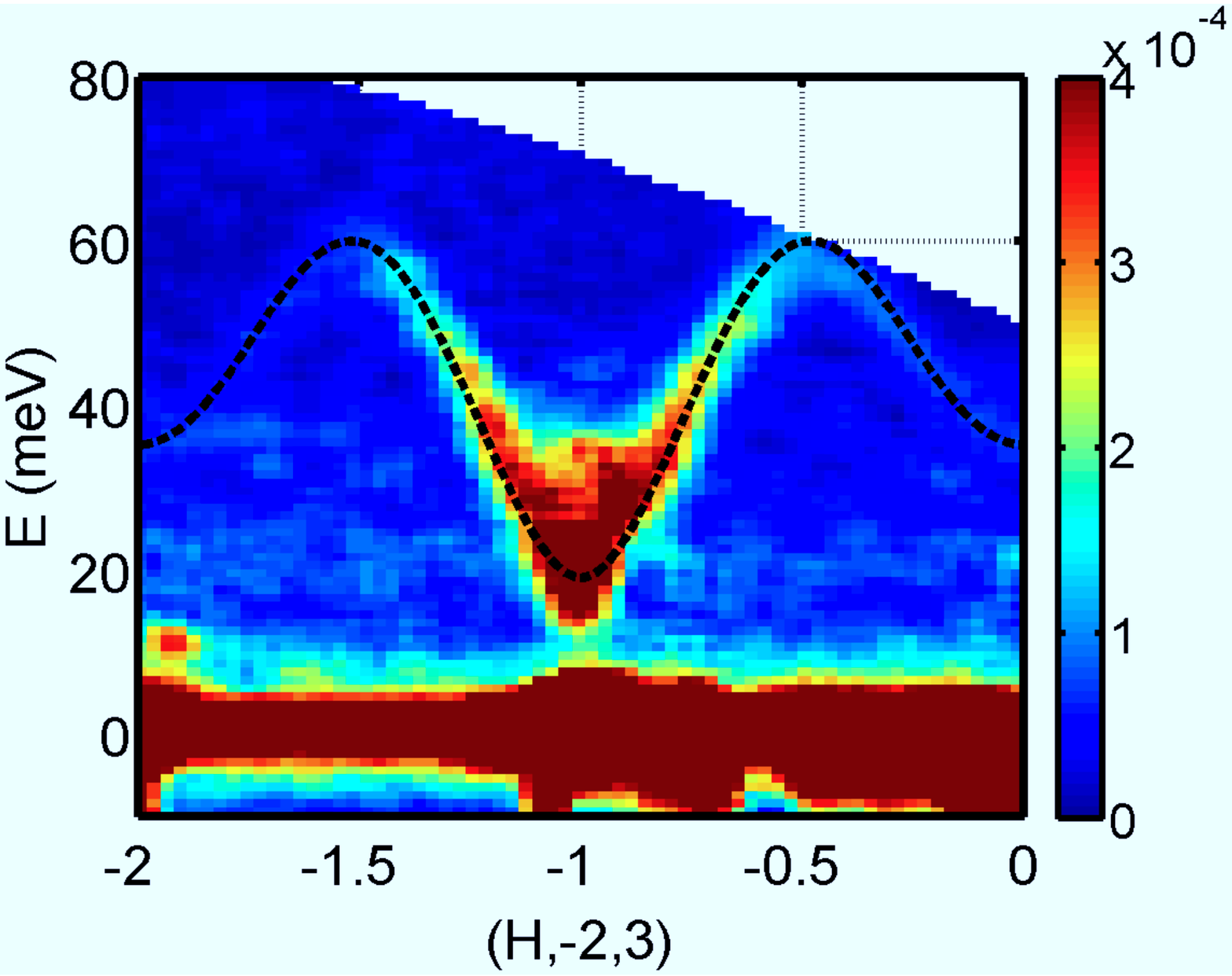} 
          \includegraphics[trim=0.7cm 6.2cm 0.7cm 6.2cm,clip=true, width=1.0\columnwidth]{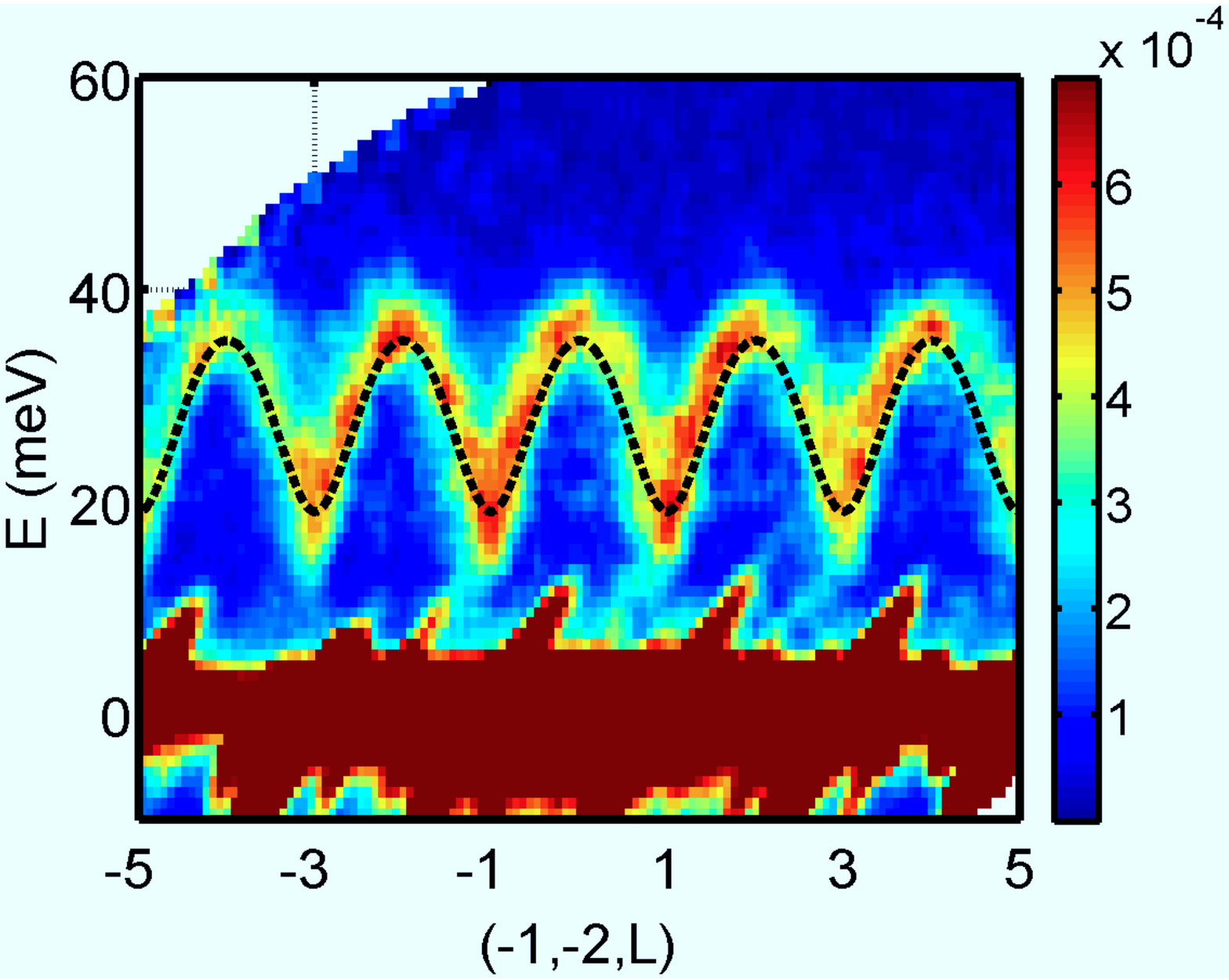} 
                   \caption{\label{Edisp} Inelastic neutron scattering measurements on single crystals of  BaMn$_2$Bi$_2$ show well defined dispersions along high symmetry directions. The intensity scale is in arbitrary units. The black dashed lines corresponds to the calculated dispersion relation using exchange interactions as described in the text.}
\end{figure}

Figure \ref{ZB} shows the energy variation and width through the spin waves at the zone boundary. The zone boundary energy is determined to be 34.2(3) meV, with a FWHM of 2.42(8) meV. With an instrument resolution at 34 meV of 1.9 meV this indicates sharp excitations, in contrast to what is observed for the Fe-122 materials, such as BaFe$_2$As$_2$ \cite{PhysRevB.84.054544}, that show diffuse scattering at the zone boundary. This distinction is likely an indication of a local moment picture being more instructive for BaMn$_2$Bi$_2$ as compared to the itinerant Fe-122 parent superconductors. 

Theoretically the spin-gap is accounted for as  a breaking of the Heisenberg spin rotation symmetry due to single ion anisotropy in the model Hamiltonian and we use this in our fitting method. We start from the Heisenberg Hamiltonian  that has proven effective for the low energy excitations in the Fe-122  parent superconducting materials such as CaFe$_2$As$_2$ and BaFe$_2$As$_2$ \cite{PhysRevLett.102.187206, PhysRevB.78.220501}:

\begin{equation}
H = \sum _{<ij>}(J_{ij}) S_i.S_j + \sum _{<i>}  {D}(S_z ^2)_i
\end{equation}
   
\noindent where $J_{ij}$ are exchange constants. We consider the $ab$-plane nearest neighbor ($J_1$) and next nearest neighbor ($J_2$) interactions as well as the interaction from the  $c$-axis nearest neighbor  ($J_C$), shown schematically in Fig.~\ref{xtal_Js}, with the spin wave dispersion given by

\begin{equation}
E(q)= \sqrt{A(q)^2-B(q)^2}
\end{equation}

\noindent In order to compare the exchange interactions with the related Mn-122 material BaMn$_2$Bi$_2$ and remain in tetragonal notation we use the  form of dispersion presented by Johnston {\it et al.}  in Ref.~\onlinecite{PhysRevB.84.094445}, with the addition of a single ion-anisotropy term ($D$) due to the observed spin-gap, with

\begin{equation}
A(q) = 2+ \frac{J_c}{J_1} -  \frac{J_2}{J_1} [2-\cos (q_xa) - \cos(q_ya)] + D 
\end{equation}

\begin{equation}
B(q) = \cos([q_x+q_y]\frac{a}{2}) + \cos([q_x-q_y]\frac{a}{2}) + \frac{J_c}{J_1}\cos(\frac{q_zc}{2})
\end{equation}

\noindent where $a$ and $c$ are the tetragonal lattice constants of BaMn$_2$Bi$_2$ and the spin-wave energy is in units of $S J_1$. This dispersion relationship is equivalent to those in Refs.~\onlinecite{PhysRevLett.102.187206, PhysRevB.78.220501}. For the case of G-type AFM with spins along the $c$-axis the exchange interactions must satisfy the constraints that $J_1$ and $J_C$ are positive and $J_1 > 2J_2$. 

Figure \ref{Edisp} shows the measured spin wave dispersions projected onto high symmetry directions in BaMn$_2$Bi$_2$. To obtain the exchange interaction values we took constant q cuts with varying energy transfer through the dispersions and fit the results with a gaussian convolved with the instrument resolution to obtain the position of the dispersions. The results were then modeled with the dispersion relationship, equation (2), along both H and L directions to find the following unique set of exchange interactions that describe the magnetic spin excitations: $SJ_1$ = 21.7(1.5), $SJ_2$ = 7.85(1.4), $SJ_C$ = 1.26(0.02), $SJ_S$ = 0.046(0.006). The exchange values are consistent with the constraints for G-type AFM ordering. Within the resolution of our measurements we do not find any change in exchange interactions between 4 K and 120 K.

Comparing the exchange interactions with those of BaMn$_2$As$_2$ reported in Ref.~\onlinecite{PhysRevB.84.094445} shows a reduction in all values in BaMn$_2$Bi$_2$, and a lowering of the top of the observed excitations from 70 meV to 55 meV. The lowering of the excitations may be anticipated from the overall reduction in magnetic ordering temperature of $\sim$200 K from BaMn$_2$As$_2$ to BaMn$_2$Bi$_2$.  The largest relative difference between the exchange interactions is the reduction of the interaction along the $c$-axis from 3 meV (BaMn$_2$As$_2$) to 1.26 meV (BaMn$_2$Bi$_2$). This may be explained in part as a consequence of the increase in the interplanar separation from 6.7 $\rm \AA$ for the MnAs layers in BaMn$_2$As$_2$ to 7.3 $\rm \AA$ for the MnBi layers in BaMn$_2$Bi$_2$ due to the $c$-lattice constant increase from 13.4149(8) $\rm \AA$ (BaMn$_2$As$_2$) to 14.687(1) $\rm \AA$ (BaMn$_2$Bi$_2$). However, the relative change of $\sim 8\%$ is the same as the reduction in the Mn-Mn interplanar distance between BaMn$_2$Bi$_2$ and BaMn$_2$As$_2$, suggesting further mechanism are important, such as the changes induced by the bismuth ion over the arsenic ion in the lattice.

Considering results for the exchange interactions in the $J_1$-$J_2$-$J_c$ model for CaFe$_2$As$_2$ \cite{PhysRevLett.101.227205,PhysRevLett.102.107003,ISI000269132100013}, BaFe$_2$As$_2$ \cite{PhysRevB.79.054526,PhysRevLett.102.107003}  and  SrFe$_2$As$_2$  \cite{PhysRevLett.101.167203, PhysRevLett.102.107003} we find reduced values for all exchange interactions in the form $SJ$ in BaMn$_2$Bi$_2$ apart from a lower $J_c$ value in BaFe$_2$As$_2$, however comparable single-ion anisotropy values. Therefore, in general, it appears the spin-waves and exchange interactions in BaMn$_2$Bi$_2$ are lower in energy compared to 122 materials, even though the magnetic ordering temperature lies between BaMn$_2$As$_2$ and the Fe-122 materials. The overall divergence of excitations energy and width, despite similarities such as the observed spin gap and structural transition in the magnetic phase, is a reflection of the differing underlying physical properties between BaMn$_2$Bi$_2$ and Fe-122 materials and further investigations via doping would be of interest to follow the evolution of the excitations in this Mn-122 material.

\section{Conclusions}

Neutron scattering measurements on single crystals have revealed BaMn$_2$Bi$_2$ forms G-type AFM at 390 K with spins aligned along the $c$-axis. We find an ordered moment of $\sim$$75\%$ compared to the expected spin-only value indicating possible hybridization and divergence from pure local moment behavior, however BaMn$_2$Bi$_2$ appears to reside much closer to the local moment limit than itinerant Fe-122 systems.  Our elastic neutron scattering suggests BaMn$_2$Bi$_2$ undergoes a subtle structural transition, similar to the Fe-122 materials but distinct from the related Mn-122 material BaMn$_2$As$_2$. A general feature of Fe-122 systems is the existence of a spin-gap as observed in inelastic neutron measurements, while the underlying relationship, if any, to superconductivity remains an open question.  Our measurements on gram sized single crystals reveal well defined spin waves and a gap of  16 meV in the low energy excitations in BaMn$_2$Bi$_2$, not seen in BaMn$_2$As$_2$, that remains unchanged in the magnetic regime from 5 K to 120 K. Indeed the spin excitations show no apparent change through the structural transition. Applying a $J_1$-$J_2$-$J_c$ Heisenberg model accounts well for the spin excitations and shows a lower energy scale compared to both BaMn$_2$As$_2$ and the Fe-122 materials. Overall our results are consistent with the postulate that the Mn-122 materials hosts intermediate properties between the local moment antiferromagnet insulating cuprate parent materials  and  itinerant antiferromagnetic Fe systems. Therefore in the context of investigating phenomena related to unconventional superconductivity BaMn$_2$Bi$_2$ appears to be well suited as a potential new bridging material.

\begin{acknowledgements}
This research at ORNL's High Flux Isotope Reactor  and Spallation Neutron Source was sponsored by the Scientific User Facilities Division, Office of Basic Energy Sciences, U.S. Department of Energy. Research was supported by the U.S. Department of Energy (DOE), Basic Energy Sciences (BES), Materials Sciences and Engineering Division (BS, AS).
\end{acknowledgements}



%

\end{document}